# Experimentos simples demostrando algunas propiedades de lentes difractivas y redes espirales

# Simple experiments demonstrating some properties of diffractive lens and spiral gratings

José J. Lunazzi [(1)], Daniel S. F. Magalhães [(1)], Maria C. I. Amon [(1)], Noemi I. R. Rivera [(1)]

1. DFMC, Instituto de Física Gleb Wataghin, Universidade Estadual de Campinas, Cidade Universitária "Zeferino Vaz" Barão Geraldo - Campinas, São Paulo, Brasil Cep 13083-970 P. Box 6165

Email de contacto: lunazzi@ifi.unicamp.br; dsouza@ifi.unicamp.br; mariaclara.amon@gmail.com; nrivera@ifi.unicamp.br

**RESUMEN:**

Introducimos algunos experimentos novedosos donde la difracción es demostrada con elementos simples: difracción de luz blanca por una moneda, construcción de una lente difractiva por holografia, propiedades de la difracción por discos digitales de información y un interesante experimento de generación de imágenes con ellos. Estos experimentos conducen a la idea de una nueva óptica enteramente difractiva, que es una tendencia en desarrollo por las facilidades de construcción, reducción de peso y otras ventajas que acarreta.

**Palabras clave:** Difracción, Óptica, Enseñanza

**ABSTRACT:**

We introduce some new experiments where light diffraction is demonstrated with simple elements: white light diffraction with a coin, construction of a diffractive lens by holography, diffraction properties in digital discs and an interesting experiment that generates images with them. This experiments lead to the idea of a new optics that is entirely diffractive, which is a trend in development for its construction facilities, weight reduction and other benefits it causes.

## 1. Introducción

En los programas de la escuela secundaria y también de la universidad la difracción es tema que se coloca después de haber desarrollado la óptica geométrica, siguiendo una línea histórica que acompaña la de los descubrimientos. Durante los siglos pasados la difracción no tenía aplicación destacada en los instrumentos ópticos, cámaras fotográficas, etc. Los hologramas son hoy elementos populares en tarjetas de acreditación etc., y recientemente fué comercializado el primer objetivo fotográfico que contiene un elemento difractivo [1]. Esto hace que el fenómeno comienze a ser cuestionado por los estudiantes y que en las primeras clases de óptica surjan preguntas sobre su princípio de funcionamiento. Nuestra propuesta es pues que los estudiantes se familiarizen con experimentos de difracción ya en sus primeros cursos, independientemente de que una teoria más completa solo les sea presentada en los cursos posteriores. Desde hace más de cinco años presentamos de esta manera una explicación de la imagen holográfica [2] en eventos de divulgación: introduciendo conjuntamente experimentos de óptica geométrica y ondulatória en un principio, para después recién llegar a la descripción del fenómeno holográfico que, por ser generalmente presentado como de óptica puramente ondulatoria, acaba ocultando sus características geométricas.

## 2. Difracción de luz blanca por un disco

El primer experimento que proponemos está inspirado en el célebre experimento de difracción de Arago [3] en el cual un disco iluminado perpendicularmente muestra una mancha luminosa en el centro de su sombra. En la discusión donde Fresnel defendía la teoría ondulatoria el matemático Poisson hizo notar que en ese caso en el centro de una sombra debería haber una mancha luminosa. La manera gráfica geométrica como era tratada la difracción nos permite fácilmente llegar a esa conclusión: para cada elemento de onda generado en el espacio que rodea al disco habrá un elemento igual y simétrico respecto del centro y así las ondas que llegaren al centro de la sombra habrán recorrido un camino de igual extensión, llegando en fase e interfiriendo constructivamente.

No sabemos exactamente con que fuente luminosa trabajó Arago, pero seguramente ha sido una de dimensiones pequeñas tal como la imagen del sol formada por un glóbulo de miel, y monocromatizada por medio de un filtro rojo [4]. Este experimento puede realizarse hoy en día fácilmente en clase: por medio de una moneda y un simple laser semiconductor con haz levemente divergente de modo que abarque un área algo mayor que la moneda, iluminándola desde unos dos metros de distancia. En una pantalla a dos o tres metros de distancia veremos exactamente en el centro de la sombra un punto luminoso, siempre centrado aunque desplazemos levemente la moneda para los lados. La divergencia del haz se consigue en aquellos laseritos que permiten desplazar la lente de colimación. Demostrar el experimento solamente con laser tiene, a pesar de su facilidad, una contraparte poco pedagógica: puede dar la idea de que el fenómeno solo ocurre con luz puntual, monocromática y coherente. Para extender en la práctica la idea al caso de luz blanca podemos usar una pequeña lámpara con filamento de 1 mm de longitud, mas no veremos la famosa mancha porque si bien la fuente tiene más intensidad que un laserito, tiene un brillo muy inferior por ser menos puntual. La mancha brillante que deberíamos encontrar tiene el tamaño y la intensidad aproximada de un disco da Airy [5] porque a la distancia en que observamos la difracción puede tratarse aproximadamente como una difracción de campo lejano, o sea, de Fraunhoffer. El diámetro es pues 2,44 $\lambda$ z/A, donde $\lambda$ es la longitud de onda, z la distancia entre el disco y la pantalla y A el diámetro del disco, una décima de milímetro aproximadamente. Podemos perfectamente considerar que el disco funciona como una lente difractiva, y que lo que tenemos en la pantalla es la imagen de la fuente. A pesar de que teóricamente la junción semiconductora que genera luz en un diodo blanco debería ser muy pequeña y por lo tanto sería una fuente muy puntual, probamos con uno, y tampoco lo logramos. Tal vez sea por causa del encapsulamiento, que podría generar aberraciones. No debemos pensar que el fenómeno no ocurre sinó que es débil. Apelamos entonces al instrumento más sensible que poseemos: el ojo. Colocandolo en el centro de la sombra vemos alrededor de la moneda un círculo blanco brillante. Así, una pequeña lámpara de tungstenio usada en paneles de automóvil es suficiente para notar el fenómeno, que no podría ser explicado por reflexión.

La figura 1 muestra la moneda que se usa como elemento difractivo, de 2 cm de diámetro, pegada a un vidrio y en un soporte simple.

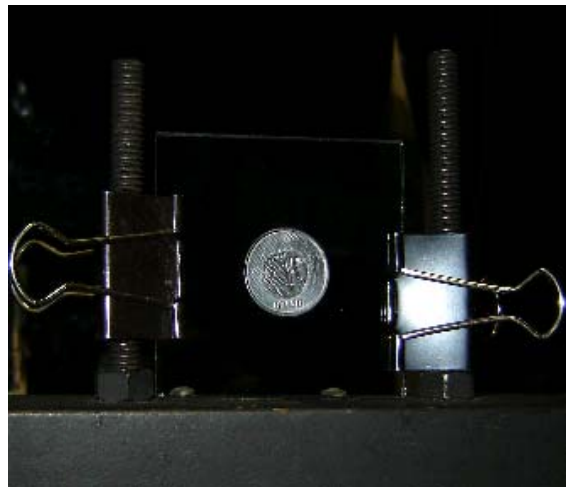

Fig. 1: Moneda pegada a un vidrio y en un soporte simple

En el experimento, la lámpara de tungstenio representa el Sol mientras que la moneda representa la Luna.

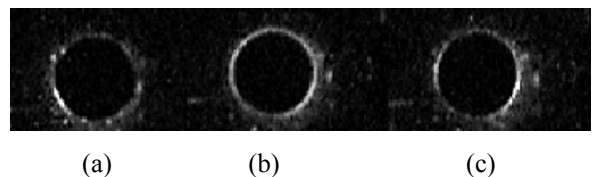

(a) (b) (c)
Fig. 2: Secuencia de imagenes del eclipse simulado

En la figura 2(a) la lámpara apenas comenzó a quedar cubierta por la parte izquierda de la moneda y consecuentemente hay una mayor luminosidad en el lado izquierdo. En la figura 2(b) la lámpara está oculta simétricamente por detrás de la moneda y se simula el momento de apogeo de un eclipse total. En la figura 2(c) tenemos la situación simétrica a la de la figura 2(a), con mayor luminosidad por el lado derecho. Nuestra secuencia de fotos simula así las tres fases de un eclipse en el caso en que la luna presenta un diámetro aparente mayor que el del sol. Muestra como la difracción puede generar un borde luminoso alrededor de la imagen del borde del disco interpuesto. En un eclipse de sol, la luna funciona como un obstáculo en el que la luz se desvía y genera un contorno. En otros términos, podemos decir que la luna podría funcionar como una lente difractiva, captando la luz del sol y permitiendo que la veamos aunque estemos bien en el centro de la región de sombra. Si buscamos los valores de los diámetros angulares de la luna y del sol veremos que son prácticamente los mismos – cerca de 0,5º. Una explicación más exacta sería que la excentricidad de la órbita de la tierra en torno del sol varía 1,67 % en torno de su media de 31' 59". La órbita de la luna alrededor de la tierra tiene una excentricidad de 0,05 y su diámetro angular máximo es de 33' 16", siendo mayor que el diámetro angular mínimo del sol (31' 28"). Es por ello que la luna parece caber perfectamente en el disco solar durante un eclipse total, tapando la parte brillante del disco y permitiendo ver un anillo brillante que sería compuesto por la corona solar, o por la difracción por el contorno de la luna, o por ambos (Fig.3) [6].

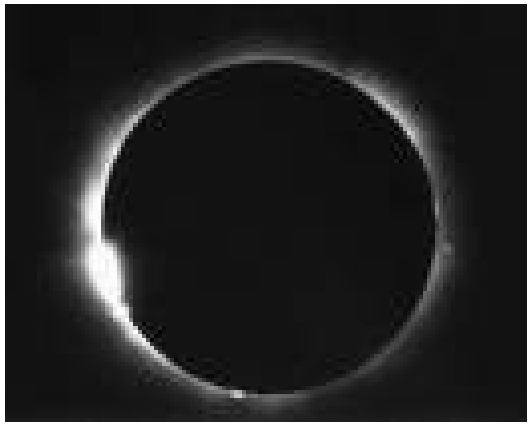

Fig. 3: Fotografía de un eclipse de sol

## 3. Construcción de una lente difractiva por holografía

Consideramos que resulta muy didáctico ofrecer a los jóvenes experimentos de difracción porque una nueva óptica se está gestando basada en elementos finos y de fácil replicación cuya utilidad crece año tras año como el telescopio espacial que ha sido propuesto y está siendo desarrollado por Hyde [7]. Además de ayudar a entender la holografía, la construcción de una lente difractiva por medio de un simple laser de diodo resulta un experimento que puede hacerse facilmente, si hay un ambiente libre de vibraciones y si se tienen dos espejos de primera superficie. Espejos así no son caros pero se pueden lograr plateando quimicante una placa de vídrio del tipo que se construye por flotación, que es el que tiene la superfície más plana, o evaporando aluminio en el caso de que se consiga quien lo haga en una cámara evaporadora. La figura 4 muestra el esquema para hacer la lente.

Se utiliza un laser de diodo (L) emitiendo en el rojo del que se ha sacado la lente colimadora. El haz resulta así de unos 5 mW de potencia, rectangular, muy divergente y limpio. Resulta preferible que la parte alargada del haz sea la que corresponde a una perpendicular al plano de la figura. En algunos modelos se puede desenroscar, otros la tienen pegada y hay que soltarla con cuidado. Se necesita que la diferencia de curvatura de los haces sea considerable para que tenga una distancia focal lo más corta posible, por ello al espejo EP1 se lo ubica luego después de la salida del haz principal y genera el haz que llamamos de objeto (HO) e incide a $45^0$ aproximadamente. EP2 es un segundo espejo y sireve para el haz de referencia (HR). Lo que caracteriza al sistema es el valor de la distancia entre él y la película: 1,7 m. Porque el haz HO ha divergido menos su intensidad sobre la película holográfica es mayor, y necesitamos atenuarla por medio de un filtro de absorción (A) que puede ser también algún filtro colorido. Intentando con varios filtros encontraremos el que deje la intensidad de los dos haces más próxima, se puede medir esto con un fotómetro de fotografia o hacer la estimativa visual, alternando la presencia de un haz o el otro sobre un papel blanco colocado en la posición de la película.

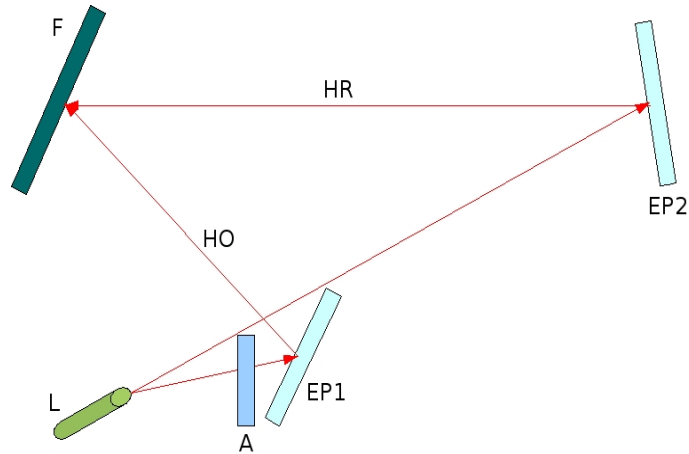

Fig.4: Esquema para hacer una lente difractiva.

La película debe sostenerse entre dos placas de vídrio de 2 mm de espesor y sujetarla por medio de prendedores para papel, apoyando el conjunto por medio de un soporte que, para ser más estable, conviene que tenga apoyo sobre tres puntos y no cuatro. El tiempo de exposición es del orden de 7 s en el que hay que cuidar de no apoyarse sobre la mesa (se podría también trabajar sobre el piso) y que no haya corrientes de aire. Se puede exponer una película [8] de hasta 10 cm x 12 cm. Revelamos [9] y consideramos que la exposición ha sido buena cuando la película queda bastante oscura, de tal modo que mirando a través a un tubo fluorescente se lo llega a ver, pero no se puede leer un texto de buen contraste impreso sobre una hoja de papel. Blanqueamos [10] y recolocamos la película entre las placas de vidrio nuevamente en su soporte retirándolo del sistema para verificar el efecto de lente. Tomamos ahora el laser recolocando su lente colimadora dejando al haz colimado o levemente divergente y lo hacemos incidir por el lado opuesto al que llegó HR. Con un poco de ajuste en las inclinaciones veremos emerger un haz en la dirección HO que irá a converger a una distancia algo mayor de lo que era la distancia al laser. Podremos entonces aproximar o alejar al laser y verificar que la lente forma una imagen que se aproxima o aleja de ella en la medida en que él lo hace.

Digamos también que si la difracción no se nota, podría ser que fuese por inestabilidad mecánica, térmica o hasta falta de coherencia del laser. En la mayoria de los laseres que hemos probado, sin embargo, hemos encontrado coherencia. Una manera de probarla sin exponer holograma consiste en colocar al haz con un poco de divergencia incidiendo perpendicularmente sobre una lámina de vidrio de unos 4 mm de espesor. Colocamos una hoja de papel inmediatamente a la salida del haz con un agujero central para que éste pase, y las reflexiones en las dos superfícies de la lámina deberán mostrarnos anillos o franjas circulares nítidas, lo que equivale a una coherencia mínima de 10 mm. Según nuestra experiencia, todo laser que tiene esa coherencia mínima comprobada posee en realidad una coherencia mayor y permite hacer hologramas. También puede comprarse un laser específico para holografia [11]. Aún habiendo coherencia, si hay inestabilidad térmica puede ser que durante la exposición el laser pase a emitir de un modo a otro, y ésto crea franjas que pueden aparecer sobre el holograma.

Una demostración bonita que puede realizarse con esa lente es usarla como pantalla holográfica. Las pantallas holográficas son el elemento más de vanguardia con el que se está desarrollando la proyección de imágenes holográficas [12, 13]. Una imagen que parece holográfica se logra fácilmente si contamos con una lámpara de filamento bien largo [14, 15]. La manera más práctica de tener un filamento así es emplear una lámpara halógena de 1.000 W montada sobre un soporte metálico cualquiera, de preferencia un perfil en forma de U que resguarda al observador de ver el filamento y usar una tensión de alimentación que sea la mitad de la nominal, de modo que con cuatro veces menos potencia el brillo es suficiente y el calor generado tolerable. La lámpara tiene un filamento de 30 cm de longitud y suele venir con sus contactos fijos a ella, de modo que lo único que hay que hacer es atornillar los cables y conectarla a la salida de un transformador. Si su tensión nominal es de 220 V la alimentaremos con 110 V. Se coloca el filamento por debajo de la lente ahora siendo pantalla de modo que, incidiendo la luz a 45$^0$, quien se posiciona frente a ella encuentra una distancia en la que ve toda la pantalla iluminada con alguno de los colores del espectro, que cambia en función de la posición vertical de observación. Hecho esto, basta colocar un objeto translúcido o una figura de alambre que veremos por detrás de la pantalla una imagen ampliada de él con paralaje horizontal continuo como si fuese una imagen holográfica. La figura 5 muestra la distribución de los elementos y ejemplifica como se obtiene cada punto imagen en lo que es una sombra, pero que el esquema de la figura 6 hecha para analizar el caso de la visión binocular demuestra el porqué de la tridimensionalidad y el paralaje de la imagen.

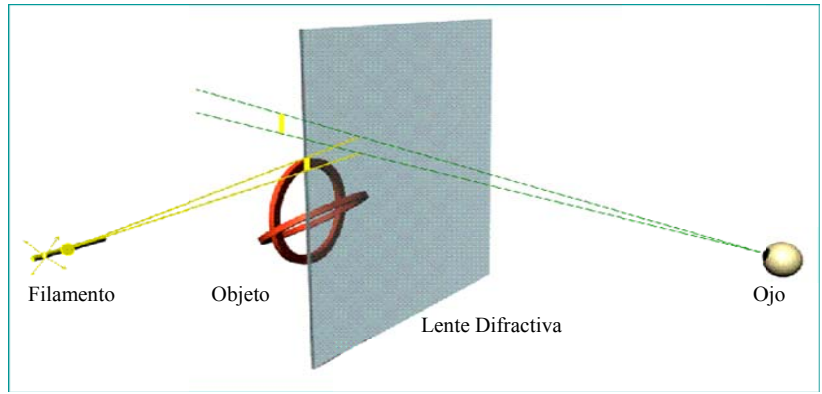

Fig.5: Distribución de los elementos y obtención de un punto imagen.

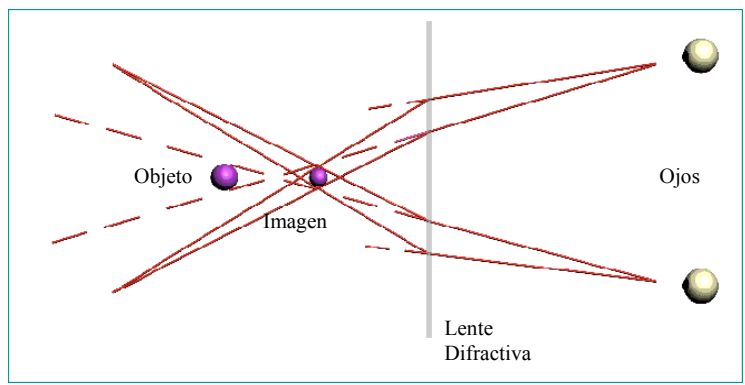

Fig. 6: La visión binocular de la imagen tridimensional.

La figura 7 muestra la fotografia de una imagen típica.

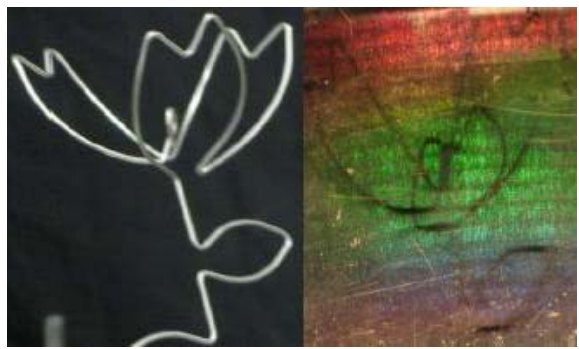

Fig. 7: A la derecha el objeto y a la izquierda una foto en blanco y negro de la imagen tridimensional.

## 4. Difracción por un CD o DVD

No hay quien no haya notado los bonitos colores que suelen producirse cuando un disco es iluminado. Esto ocurre sobre todo cuando hay lámparas poco extensas. A partir de un elemento tan común es posible demostrar sus propiedades difractivas con un montaje muy simple. Para hacer el experimento podemos usar una pequeña lámpara fluorescente porque tiene longitudes de onda discretas, ubicándola a 1-2 m de distancia y de modo que presente un área pequeña, eventualmente, cubriéndola parcialmente. Cuando un observador pone su ojo a una distancia de unos pocos centímetros él ve las líneas de emisión de la lámpara. La figura 8 es una fotografía tirada con una cámara en esa posición.

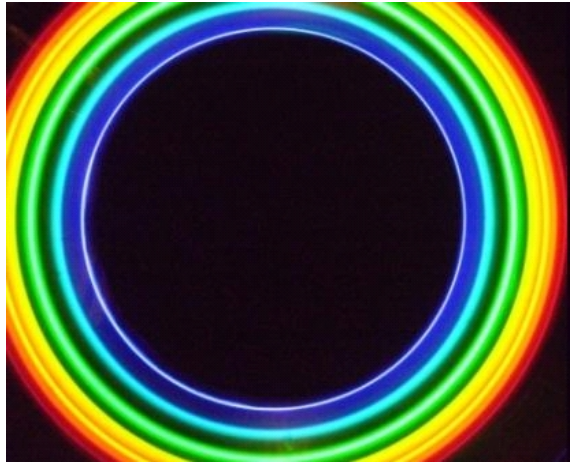

Fig.8: La cámara en la posición z= 9,2±0,1 cm. Los colores del espectro de la lámpara pueden ser vistos.

Así, un elemento simple como un CD o DVD puede ser usado como ejemplo de espectrometría hasta en una sala de aula puesto que el experimento puede realizarse aunque haya luz ambiente.

Para conseguir un CD o DVD por transmisión, podemos recurrir a la caja de 50 unidades donde podemos encontrar un CD transparente. Otra opción es retirar el borde con una tijera, de esta manera la película reflectora saldrá.

Usualmente un CD tiene aproximadamente 650 líneas/mm y el DVD 1200 líneas/mm.

Para adentrarnos en el fenómeno de difracción de un CD describiremos la teoría que lo justifica.

Considere el CD o DVD como un elemento con una ranura en formato espiral (como podemos observar en los antíguos discos de vinil) y con perfil dado por una función periódica cualquiera. Esa función puede ser expandida en série de Fourier como:

$$t(r) = \sum_{n=-\infty}^{\infty} c_n \exp[j.2.\pi.{r}/{r_0}] \qquad (1)$$

donde $c_n$ son constantes complexas y $r_0$ es la distancia radial entre vueltas adyacentes.

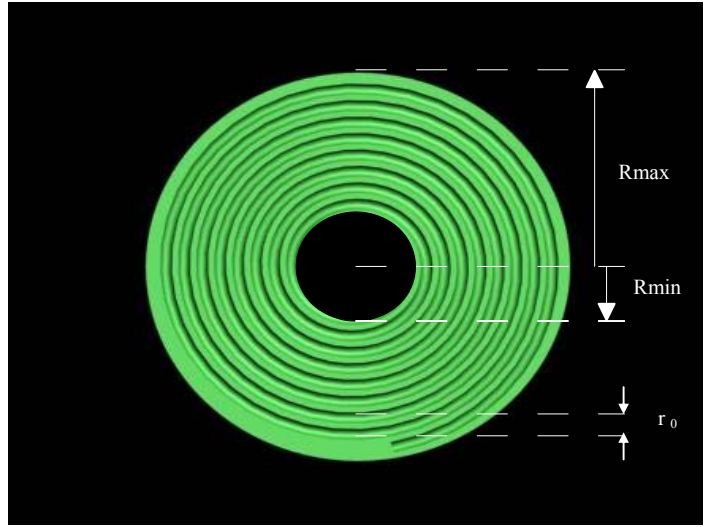

Fig.9: Figura esquemática representativa de un CD usando el modelo espiralado.

Haciendo la substitución $2\pi r/r_0 \rightarrow 2\pi r/r_0 - \theta$ obtenemos una expresión para el perfil espiral válida para todos los valores del ángulo polar. Una función de transferencia puede ser:

$$t_A(r,\theta) = \left[ circ\left(\frac{r}{R_{max}}\right) - circ\left(\frac{r}{R_{min}}\right) \right] \times \sum_{n=-\infty}^{\infty} c_n e^{j.n.\left(2.\pi\frac{r}{r_0} - \theta\right)} \quad (2)$$

Es conocido [16] que haciendo el cálculo de la difracción en la aproximación de Fresnel (paraxial) llegamos a una expresión para el campo igual a:

$$E_n(r',\theta',z) \approx n E_0 c_{-n} \pi \cdot j^{-(n+1)} e^{jkz} \times e^{j\frac{\pi}{4}} \sqrt{\frac{z\lambda}{4r_0^2}} e^{jn\theta'} e^{-j\pi.n^2\frac{z\lambda}{r_0^2}} J_n\left(\frac{2\pi n r'}{r_0}\right) \quad (3)$$

Donde r' y θ' son las coordenadas del campo $E_n$ en el plano de observación; z es la distancia entre el CD (o DVD) y el plano de observación; $E_0$ es el módulo del campo colimado incidente; $r_0$ es la distancia radial entre vueltas adyacentes en el CD (o DVD); $c_{-n}$ es la enésima componente de la función que caracteriza el perfil de un CD (ejemplo: una función de Bessel de enésima orden); n es un orden de difracción, en nuestro caso n=1; λ es la longitud de onda de la luz incidente; $R_{min}$ y $R_{max}$ son el mínimo y el máximo radio del CD (o DVD), respectivamente.

La expresión (3) es válida para

$$(r_0/n\lambda) R_{min} < z < (r_0/n\lambda) R_{max} \quad (4)$$

y ordenes de grandeza menor fuera de esta región. Podemos notar que para z=0 la ecuación (3) no es válida, pues en la expresión (4) ro debe siempre ser diferente de zero.

La figura 10 muestra las líneas formadas por el haz para dos órdenes de difracción (n=1 y n=3).

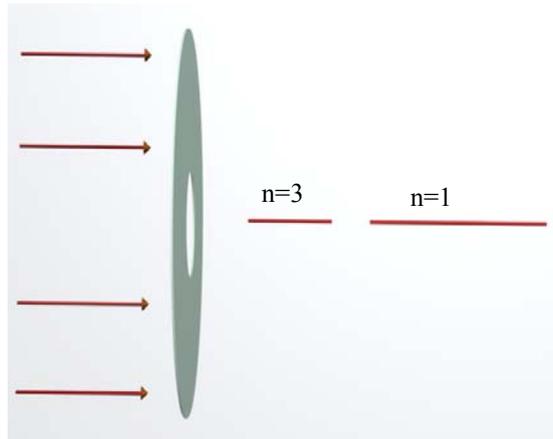

Fig. 10: El haz incidente colimado forma dos líneas de longitud.

Con una fuente de luz blanca esperamos que cada longitud de onda esté en una región con z diferente, pues como vemos en (4) hay una dependencia con la longitud de onda. La figura 11 muestra el resultado esperado para iluminación policromática. Nótese que hay uma región donde llegan todas las longitudes de onda, o sea, el segmento blanco.

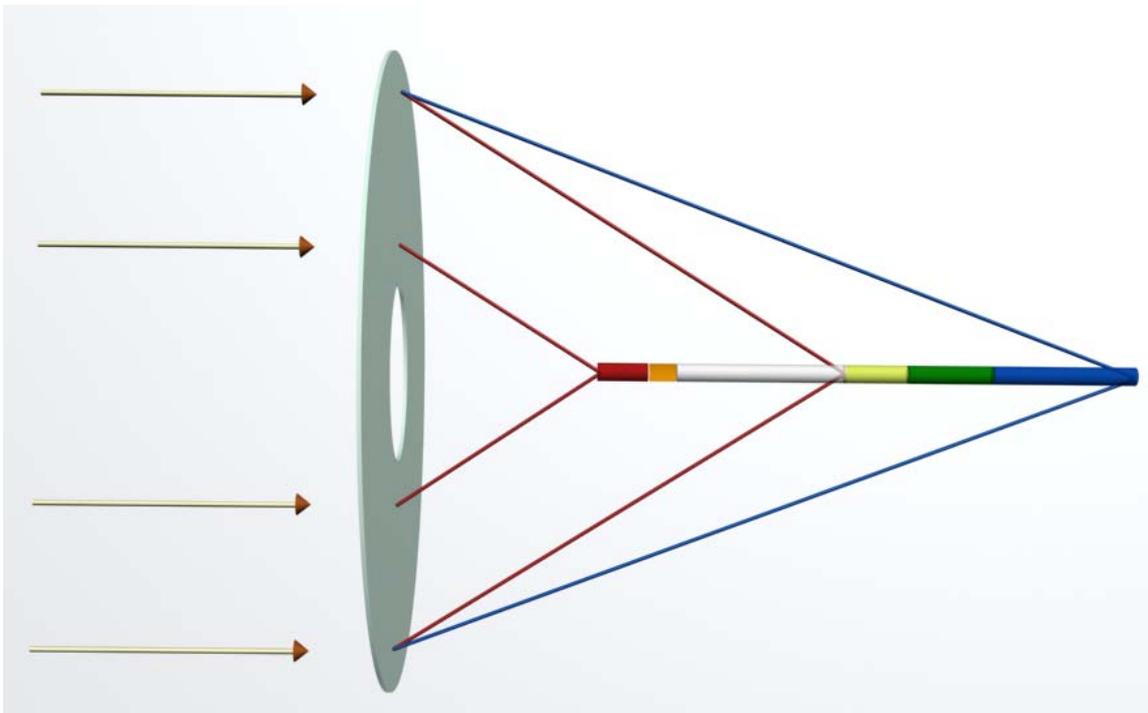

Fig. 11: El haz incidente colimado forma un conjunto de líneas superpuestas.

Entendemos ahora que el ojo del observador en nuestro experimento debe estar sobre la parte blanca de la línea z donde la relación (4) es cumplida.

## 5. Realización de imágenes en luz blanca por medio de óptica puramente difractiva

Ya familiarizados con la conducta óptica de los discos digitales podemos ahora encarar un experimento enteramente novedoso y fascinante por lo sencillo: la realización de una imagen usando un disco como primer componente y otro como segundo. El fenómeno básico es el de la obtención de imágenes virtuales cuando un primer elemento tiene el doble del número de líneas por milímetro que un segundo. Puede encontrarse en ese caso una distancia entre los elementos difractivos que permita que un objeto se vea desde el segundo, como si fuese un periscopio difractivo. Casualmente, un disco DVD tiene el doble del número de líneas por milímetro que un CD. Puede hacerse con los elementos como se obtiene comunmente, o sea, con capa reflectora, pero resulta más fácil de alinear y de interpretar usando los elementos sin la capa reflectora. La capa reflectora de un DVD puede retirarse, pero en caso de no conseguirlo podemos igualmente partir de un primer disco del tipo CD que si hemos conseguido sin la capa reflectora. La equivalencia de los dos elementos viene de que en el caso del CD usaremos la segunda orden de difracción. Utilizaremos solamente una región pequeña del primer disco para recibir la luz del objeto, y otra semejante simétricamente en el segundo. La foto (figura 12) nos muestra como se han de disponer los dos discos, separados por una distancia de 8,5 cm y el objeto con la lámpara al lado. La distancia entre el objeto y el primer disco es de 24,5±0,1 cm y hay una relación de distancia entre discos para cada distancia del objeto que nuestro cálculo numérico no permite determinar teóricamente como una fórmula general. El observador mira próximo al segundo elemento.

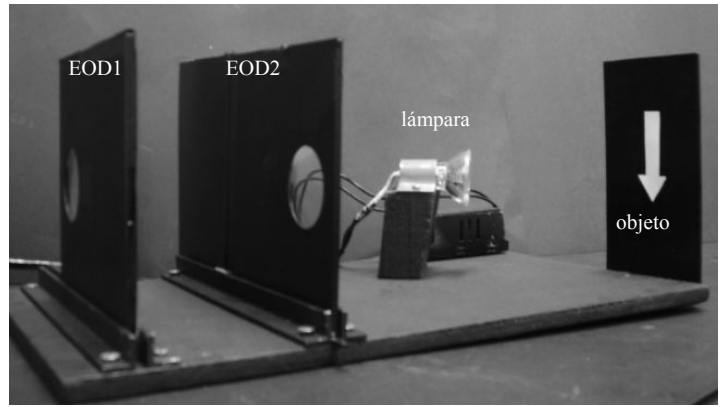

Fig 12. Foto de los elementos difractivos posicionados.

El fenómeno [17] utiliza luz doblemente difractada, como la eficiencia de cada elemento es del orden del 10%, solo 1% puede ser aprovechado y es necesario tener un objeto fuertemente iluminado. Una lámpara halógena dicroica de 50 W es el elemento adecuado. Utilizamos un objeto en forma de flecha blanca sobre fondo negro para tener el máximo contraste. La figura 13 nos muestra la foto del objeto en forma de flecha y a su lado la imagen del mismo, invertida y con alguna distorsión.

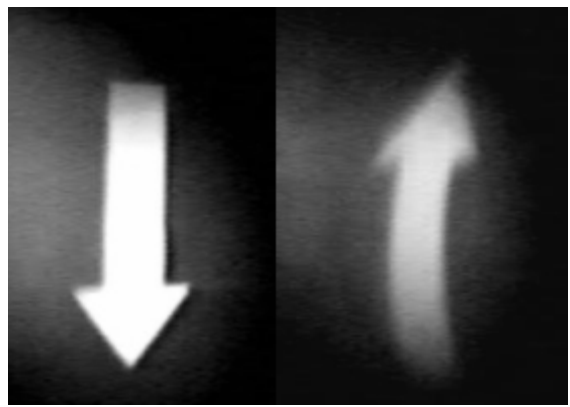

Fig.13 izquierda: objeto. Derecha: imagen

Por ser los elementos planos y transparentes, el resultado de la imagen invertida sorprende. Recordemos que por refracción una imagen invertida se obtiene con una lente convergente, ¿pero como asociar este experimento al caso de una lente convergente? Podemos explicar el fenómeno pero no encontrar la relación, veamos: la Fig. 14 muestra el porqué de la inversión de la imagen al considerar la dirección de difracción en cada parte del elemento circular siendo definida por una red simple que corresponde a la tangente de los surcos en el punto de incidencia.

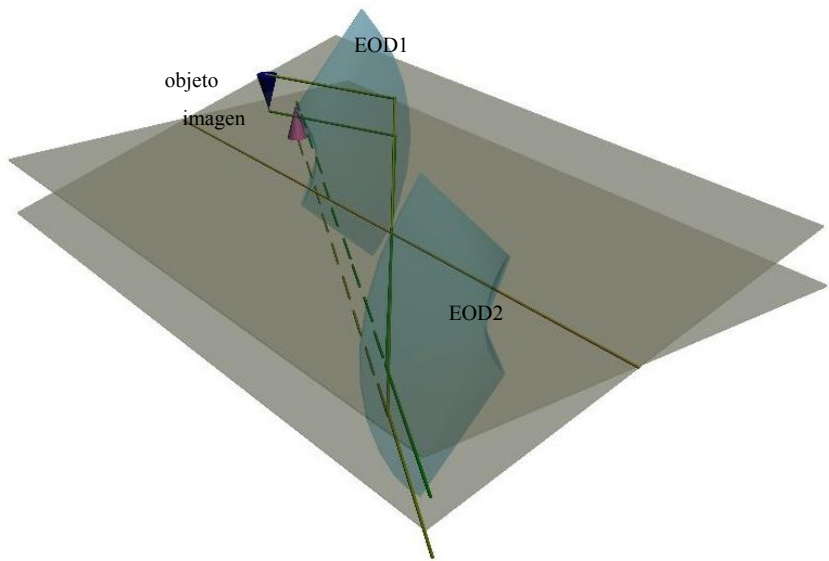

Fig. 14: camino de los rayos que salen del objeto y forman la imagen

Los rayos que inciden alto son desviados hacia abajo, los que inciden abajo lo son hacia arriba. La figura 15 muestra el esquema de rayos para el sistema de doble difracción utilizando redes espirales de una misma frecuencia espacial. Un objeto de luz blanca es colocado frente al primer elemento óptico difractivo EOD1, una línea central imaginaria pasa por el centro de curvatura de los elementos difractivos. La segunda orden de difracción corta a la línea central y se dirige a una cierta distancia al segundo elemento óptico difrativo EOD2.

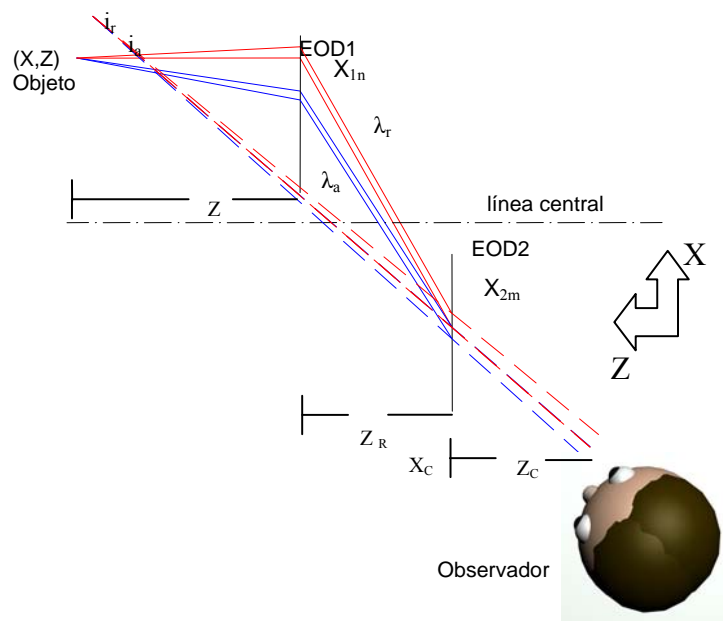

Fig. 15. Esquema de trazado de rayos para la formación de la imagen.

La dirección del desvío de la segunda difracción es opuesta a la de la primera difracción, la imagen vista por el observador corresponde a una imagen virtual del objeto.

El proceso de difracción es resuelto a través de la ecuación de la red y el esquema del proceso de imagen es mostrado en la Fig.15 donde X y Z son las coordenadas del objeto. $X_{1n}$ es la coordenada de incidencia en el primer elemento difractivo EOD1, $X_{2m}$ es la coordenada de incidencia del segundo elemento difractivo EOD2 y $Z_R$ es la distancia entre ambos elementos. λnm es la longitud de onda de cada rayo difractado en $X_{1n}$ y los rayos difractados en $X_{2m}$. Ejemplo de estos valores de longitud de onda son dados en la Fig. 15 por medio de $\lambda_r$ e $\lambda_a$, donde $\lambda_r > \lambda_a$. La primera orden de difracción en el primer elemento EOD1 esta regida por la ecuación (5).

$$\operatorname{sen}\theta_i - \operatorname{sen}\theta_d = 2\lambda\nu \tag{5}$$

donde $2\nu$ es la frecuencia de líneas en la primera difracción. Según el esquema de trazado de rayos de la Figura 15, la ecuación (5) se puede expresar en coordenadas cartesianas como:

$$\frac{X_{1n}-X}{\sqrt{X_{1n}^2+Z^2}} + \frac{X_{1n}-X_{2m}}{\sqrt{(X_{1n}-X_{2m})^2+Z_R^2}} = 2\lambda_{nm}\nu \tag{6}$$

Análogamente, el primer orden de difracción en el segundo elemento EOD2 se puede expresar como:

$$-\frac{X_{1n}-X_{2m}}{\sqrt{(X_{1n}-X_{2m})^2+Z^2}} + \frac{X_{2m}-X_C}{\sqrt{(X_{2m}-X_C)^2+Z_C^2}} = -\lambda_{nm}\nu \tag{7}$$

donde $X_C$, $Z_C$ son coordenadas de la posición del observador. Destacamos que en este caso no entra el factor 2, $\nu$ es ahora exactamente la frecuencia de líneas en la segunda difracción.

Del sistema de equaciones (6) - (7) podemos obtener os valores $X_{1n}$ e $X_{2m}$. Con los valores de $X_{1n}$ y $X_{2m}$ y la equación (8), para cada longitud de onda $\lambda_a$ y $\lambda_v$ encontramos la posición de un punto imagen (Xi, Zi) de un ponto objeto (X, Z). $Xi_a$ para la menor longitud de onda y $Xi_v$ para la mayor longitud de onda.

$$-\frac{X_{1n}-X_{2m}}{\sqrt{(X_{1n}-X_{2m})^2+Z_R^2}} + \frac{X_{2m}-Xi_{a,v}}{\sqrt{(X_{2m}-Xi_{a,v})^2+Zi_{a,v}^2}} = -\lambda_{nm}\nu \tag{8}$$

Todo este cálculo sirve para probar que el fenómeno puede ser entendido, pero podemos guardarlo solo para los más interesados, para los demás basta mostrar un esquema básico de trazado de rayos y explicar que la imagen ocurre porque la descomposición de la luz en el primer elemento es compensada en el segundo.

## 6. Conclusiones

Hemos dado un conjunto de elementos adicionales a los ya clásicos experimentos de difracción con los que esperamos que los estudiantes de hoy puedan familiarizarse con este fenómeno y hagan en el futuro aportes creativos.